\documentstyle[11pt,a4,ere99]{article}
\input{epsf}
\begin{document}

\title{Singularity-free orthogonally-transitive cylindrical
spacetimes}

\author{L. Fern\'andez-Jambrina}

\address{Departamento de Ense\~nanzas B\'asicas de la Ingenier\'{\i}a 
Naval, E.T.S.I. Navales\\
E-28040-Madrid, Spain\\E-mail: {\em {\tt lfernandez@etsin.upm.es}}}


\maketitle\abstracts{
In this talk  a previous theorem on geodesic completeness of
diagonal cylindrical
spacetimes will be generalized to cope with the nondiagonal case. 
A
sufficient condition for
such spacetimes to be causally geodesically complete will be given.}

\section{Introduction}

This talk will deal to a certain extent with the natural continuation of 
some work in progress that was introduced last year at the Spanish 
Relativity Meeting in Salamanca \cite{ere}${}^,$ \cite{manolo} and that has recently been 
published \cite{nondiag}.

The reason for investigating geodesic completeness of Lorentzian 
manifolds is twofold. From the mathematical point of view, there is no 
analogue for the Hopf-Rinow theorem that characterizes geodesic 
completeness of Riemannian manifolds in terms of metric completeness. 
Since the Lorentzian metric does not determine a metric structure, 
just a causal structure, 
this possibility is hindered.

From the physical point of view, the existence of inhomogeneous 
cosmological models that are causally geodesically complete and, 
therefore, singularity-free, and whose role in Cosmology is yet to be 
determined, induces us to try to characterize when such behaviour 
arises.

Of course, there are well-known results, the singularity theorems 
\cite{HE}${}^,$ \cite{Beem} due 
to Hawking, Penrose, Tipler \dots, that provide sufficient conditions 
for a Lorentzian manifold to be incomplete. Furthermore, they discern 
whether there is a Big Bang, Big Crunch or geodesic imprisonment 
singularity. 

On the contrary, sufficient conditions for a spacetime to be causally 
geodesically complete are not so easily come across in the literature 
\cite{miguel}${}^,$ \cite{manolo}${}^,$ \cite{nondiag}.

Since such nonsingular cosmological models have arisen within the 
framework of inhomogeneous 
cosmologies, I shall devote my talk  to  orthogonally-transitive 
$G_{2}$ cylindrical spacetimes.

\section{Geodesic equations for an Abelian orthogonally transitive $G_{2}$ 
metric.}

I shall introduce the metric in isothermal coordinates $t,r$. The 
other coordinates are cyclic and therefore do not appear explicitly in 
the expression,
\begin{eqnarray}
ds^2=e^{2\,g(t,r)}\left\{-dt^2+dr^2\right\}+\rho^2(t,r)e^{2\,f(t,r)}d\phi^2
+e^{-2\,f(t,r)}\{dz+A(r,t)\,d\phi\}^2.\label{metric}
\end{eqnarray}

The coordinates have the usual range and we shall assume from the 
beginning that every metric function is $C^2$.

For the cylindrical symmetry, I shall take the usual definition. The 
axis is located where the norm, $\Delta$, 
\begin{equation}
\Delta=g(\xi,\xi)=\rho^2(t,r)e^{2\,f(t,r)}+e^{-2\,f(t,r)}\,A^2(r,t),
\end{equation}
of the axial Killing field 
vanishes and we shall require the following behaviour for it in the 
neighbourhood of the axis,

\begin{equation}
\lim_{r\to
0}\frac{g({\rm grad}\,\Delta,{\rm grad}\,\Delta)}{4\,\Delta}=
1,\label{reg} \end{equation}
assuming that it is located on the locus $r=0$, where it can be set 
by performing a suitable coordinate transformation.

Now all we have to do is to check the behaviour of the geodesic 
equations. They form a system of four quasilinear second-order 
ordinary differential equations plus an additional first-order 
cuadratic differential equation that determines the geodesic 
parametrization up to an affinity,

\begin{eqnarray} 
\ddot x^i+\Gamma^i_{jk}\dot x^j\dot x^k = 0,\nonumber\\
g(\dot x,\dot x)=\left\{
\begin{array}{c}
    -1  \\
    0  \\
    1.
\end{array}
\right.\end{eqnarray}

The existence of isometries leads to integration of two equations, 
thereby reducing the order of the system. We have two constants of 
geodesic motion, 

\begin{eqnarray}
L=e^{2\,f(t,r)}\,\rho^{2}(t,r)\,\dot\phi+
e^{-2\,f(t,r)}\,A(t,r)\,\{\dot z+A(t,r)\,\dot \phi\},
\end{eqnarray}

\begin{eqnarray}
P=e^{-2\,f(t,r)}\{\dot z+A(t,r)\,\dot \phi\},
\end{eqnarray}
respectively the angular momentum around the axis and 
the linear momentum along the axis.

There is another constant, $\delta$, that takes the value $-1,0,1$ for 
respectively spacelike, lighlike and timelike geodesics,
\begin{eqnarray}
\delta=e^{2\,g(t,r)}\left\{\dot
t^2-\dot  r^2\right\}-
\{L-P\,A(t,r)\}^2\rho^{-2}(t,r)e^{-2\,f(t,r)}-P^2e^{2\,f(t,r)}.\label{delta}
\end{eqnarray}

As a matter of convenience, we shall denote by $\Lambda$ the `effective' 
angular momentum, 
\begin{eqnarray}
\Lambda(t,r)=L-P\,A(t,r).
\end{eqnarray}

The other equations can be written in a compact form as, 
\begin{eqnarray}
\{e^{2\,g(t,r)}\dot t\}^{\cdot} -
e^{-2g(t,r)}F(t,r)F_t(t,r)=0,
\label{eeq1}
\end{eqnarray}

\begin{eqnarray}
\{e^{2\,g(t,r)}\dot
r\}^{\cdot}+e^{-2g(t,r)}F(t,r)F_r(t,r)=0,\label{eeq2}
\end{eqnarray}

\begin{eqnarray}
F(t,r)=e^{g(t,r)}\sqrt{\delta+P^2e^{2f(t,r)}+
\Lambda^2(t,r)\frac{e^{-2f(t,r)}}{\rho^{2}(t,r)}},
\end{eqnarray}
that suggests 
a parametrization and reduction of the differential system to a first 
order one,

\begin{eqnarray}
\dot
t=\pm e^{-2g(t,r)}F(t,r)\cosh\xi(t,r),
\end{eqnarray}

\begin{eqnarray}
\dot
r=e^{-2g(t,r)}F(t,r)\sinh\xi(t,r),
\end{eqnarray}

\begin{eqnarray}
\dot\xi(t,r)=-
e^{-2g(t,r)}\left\{\pm F_t(t,r)\sinh\xi(t,r)+F_r(t,r)\cosh\xi(t,r)\right\},
\end{eqnarray}
where the upper sign is meant for future-pointing geodesics and the 
lower one, for past-pointing geodesics.

The phylosophy of the work is just preventing the coordinates $t,r$ 
from tending to infinity, and thus leaving the spacetime, at a finite 
value of the affine parameter. 

The results are quoted in the next section. Proofs can be found in 
\cite{nondiag}.

\section{Results}

The results can be summarized in two theorems, one for future-pointing 
and one for past-pointing geodesics,

\begin{description}
\item[Theorem 1:] An orthogonally transitive cylindrical spacetime endowed with a metric whose
local expression in terms of $C^2$ metric functions $f,g,A,\rho$ is given by (\ref{metric})
such that the axis is located at $r=0$ has complete future causal geodesics if the following set
of conditions is fulfilled:

\begin{enumerate}
\item For large values of $t$ and increasing $r$, 
\begin{enumerate}
\item \label{Mxi1}$\left\{
\begin{array}{l}g_u\ge 0\\
h_u\ge 0\\
q_u\ge 0,\end{array}\right.$
\item \label{Mxi2} Either $\left\{
\begin{array}{lcl}{g_r}\ge 0&\textrm{or}& 
|g_r|\stackrel{\displaystyle{<}}{\sim} g_u\\h_r
\ge 0&\textrm{or}& |h_r|\stackrel{\displaystyle{<}}{\sim} h_u\\q_r\ge 0&\textrm{or}& |q_r|\stackrel{\displaystyle{<}}{\sim}
q_u.\end{array}\right.$
\end{enumerate}
\item For $L\neq0$ and large values of $t$ and decreasing $r$, 
\begin{enumerate}
\item \label{mxi1}$\delta\,g_v+P^2e^{2f}\,
q_v+\Lambda^2\frac{e^{-2f}}{\rho^2}h_v\ge 0$
\item \label{mxi2} Either $\delta g_r+P^2e^{2f}\,
q_r+\Lambda^2\frac{e^{-2f}}{\rho^2}\,h_r\le 0$ {or} $\delta g_r+P^2e^{2f}\,
q_r+\Lambda^2\frac{e^{-2f}}{\rho^2}\,h_r\stackrel{\displaystyle{<}}{\sim} \delta g_v+P^2e^{2f}\,
q_v+\Lambda^2\frac{e^{-2f}}{\rho^2}\,h_v.$
\end{enumerate}
\item \label{tt} For large values of the time coordinate  $t$, constants 
$a,b$ exist such that, 
\begin{displaymath}\left.\begin{array}{c}2\,g(t,r)\\g(t,r)+f(t,r)+\ln\rho-\ln|\Lambda|\\
g(t,r)-f(t,r)\end{array}\right\}\ge-\ln|t+a|+b.\end{displaymath}

\item \label{ax} The limit $\displaystyle\lim_{r\to 0}\frac{A}{\rho}$ exists.
\end{enumerate}

\end{description}

\begin{description}
\item[Theorem 2:] An orthogonally transitive cylindrical spacetime endowed with
a metric whose local expression in terms of $C^2$ metric functions $f,g,A,\rho$
is given by (\ref{metric}) such that the axis is located at $r=0$ has complete
past causal geodesics if the following set of conditions is fulfilled:

\begin{enumerate}
\item For small values of $t$ and increasing $r$, 
\begin{enumerate}
\item $\left\{
\begin{array}{l}g_v\le 0\\
h_v\le 0\\
q_v\le 0,\end{array}\right.$
\item  Either $\left\{
\begin{array}{lcl}{g_r}\ge 0&\textrm{or}& |g_r|\stackrel{\displaystyle{<}}{\sim} -g_v\\h_r
\ge 0&\textrm{or}& |h_r|\stackrel{<}{~} -h_v\\q_r\ge 0&\textrm{or}& |q_r|\stackrel{\displaystyle{<}}{\sim}
-q_v.\end{array}\right.$
\end{enumerate}
\item For $L\neq0$ and small values of $t$ and decreasing $r$, 
\begin{enumerate}
\item $\delta\,g_u+P^2e^{2f}\,
q_u+\Lambda^2\frac{e^{-2f}}{\rho^2}h_u\le 0$
\item Either $\delta g_r+P^2e^{2f}\,
q_r+\Lambda^2\frac{e^{-2f}}{\rho^2}\,h_r\le 0$ {or} $\delta g_r+P^2e^{2f}\,
q_r+\Lambda^2\frac{e^{-2f}}{\rho^2}\,h_r\stackrel{\displaystyle{<}}{\sim} |\delta g_u+P^2e^{2f}\,
q_u+\Lambda^2\frac{e^{-2f}}{\rho^2}\,h_u|.$
\end{enumerate}
\item For small values of the time coordinate  $t$, constants $a,b$ exist 
such that, 
\begin{displaymath}
    \left.\begin{array}{c}2\,g(t,r)\\g(t,r)+f(t,r)+\ln\rho-\ln|\Lambda|\\
    g(t,r)-f(t,r)\end{array}\right\}\ge-\ln|t+a|+b.
\end{displaymath}

\item The limit $\displaystyle\lim_{r\to 0}\frac{A}{\rho}$ exists.
\end{enumerate}

\end{description}

The coordinates $u,v$ are respectively the usual outgoing and ingoing 
radial null coordinates. Attention needs be paid to the last 
condition of the theorems, that just states the fact that the 
geometry in the vicinity of the axis cannot be determined by the function in the metric
related to non-diagonality, $A$. This is the main novelty 
in the non-diagonal case. 

The results for past-pointing geodesics are obtained from the ones 
for future-pointing geodesics just exchanging some signs and the 
coordinates $u,v$.

The conditions of the theorems look rather lengthy, but are not 
difficult to check.

\section{Nonsingular cosmological models}

In order to check whether these theorems are too restrictive 
sufficient conditions, we shall apply them to all non-diagonal 
cylindrical perfect fluid cosmological models that are known to be 
singularity-free in the literature. The diagonal ones were already 
checked \cite{manolo}.

\begin{enumerate}
\item Mars \cite{Diag}: It is the first known nonsingular
nondiagonal cylindrical cosmological model in the literature.

\begin{eqnarray}
&g(t,r)=\frac{1}{2}\ln\cosh(2\,a\,t)+\frac{1}{2}\alpha\,
a^2r^2,\nonumber\\& f(t,r)=\frac{1}{2}\ln\cosh(2\,a\,t),\nonumber\\& 
\rho(t,r)=r,\nonumber\\& A(t,r)=a\,r^2,
\end{eqnarray}
where $a$ is a constant and $\alpha>1$. If $\alpha=1$ the pressure and the
density of the fluid vanish.

\item Griffiths-Bi\v{c}ak \cite{Jerry}: The previous model is comprised in
this one for $c=0$ after a redefinition of constants. The metric functions can
be written as,

\begin{eqnarray}
&g(t,r)=\frac{1}{2}\ln\cosh(2\,a\,t)+\frac{1}{2}
a^2r^2+\frac{1}{2}\Omega(t,r),\nonumber\\&
f(t,r)=\frac{1}{2}\ln\cosh(2\,a\,t),\nonumber\\  &\rho(t,r)=r,\qquad
A(t,r)=a\,r^2,
\end{eqnarray}
where $\Omega$ is a function that is obtained from a solution, $\sigma$, of the
wave equation,
\begin{eqnarray}
&\Omega_r=r(\sigma_t^2+\sigma_r^2),\qquad \Omega_t=2r\,\sigma_t
\sigma_r,\nonumber\\
&\sigma(t,r)=bt+\sqrt{2}{c}\sqrt{\frac{\sqrt{(\alpha^2+r^2-t^2)^2
+4\alpha^2t^2}+\alpha^2+r^2-t^2}{(\alpha^2+r^2-t^2)^2
+4\alpha^2t^2}}.
\end{eqnarray}

\end{enumerate}

Both of them are stiff perfect fluids and can be checked to fulfil 
the conditions of the theorems.

\vspace*{-2pt}
\section*{Acknowledgments}
The present work has been supported by Direcci\'on
General de Ense\~nanza Superior Project PB95-0371. The author wish to thank
 F. J. Chinea.,  L. M. Gonz\'alez-Romero, F. Navarro and M. J. Pareja for valuable discussions
.

\end{document}